# A local perspective on conjugation of double bonds in acyclic polyenes

Viktorija Gineityte

Institute of Theoretical Physics and Astronomy, Vilnius University, Sauletekio al. 3, LT-10257, Vilnius, Lithuania
Email: viktorija.gineityte@tfai.vu.lt

**Abstract**
The study is devoted to elaboration of an alternative image of conjugation in acyclic polyenes as a weak and essentially local delocalization of initially-localized pairs of $\pi$-electrons ascribed to individual double (C=C) bonds (instead of formation of a completely delocalized $\pi$-electron system as usual). To this end, polyenes are modelled as sets of weakly-interacting formally-double (C=C) bonds, where the single (C–C) bonds represent the interaction between the former and are treated as a perturbation. Mathematically, the above-formulated goal is realized by means of a particular version of the non-canonical method of molecular orbitals (MOs) based on the Brillouin theorem and yielding the expressions both for total energies and for non-canonical (localized) MOs (NCMOs) directly without any reference to usual (canonical) MOs. In addition, total energies and NCMOs are interrelated explicitly in the approach applied, viz. the former are representable via the so-called delocalization coefficients of the latter. Adaptation of these general results to the above-specified model of polyene yields coincidence between the conjugation energy (CE) and the total delocalization energy of all pairs of $\pi$-electrons contained. Moreover, a local relation follows between constitution of the nearest environment of a certain C=C bond, delocalization pattern of the respective pair of $\pi$-electrons and contribution of just this pair to the total CE of the given polyene. As a result, different stabilities of distinct polyenes (e.g. of isomers) prove to be accompanied by variable extents of delocalization of separate pairs of $\pi$-electrons. Linear and cross-conjugated polyene chains are comparatively analysed. Delocalization conditions are found to be substantially better in the former case due to participation of all C=C bonds in three-membered conjugated paths. In addition, the distinction concerned is especially striking in the middle areas of these chains. Enhancement (deterioration) of delocalization conditions with elongation of a linear (cross-conjugated) chain also is among the conclusions. For polyenes of irregular constitution, the extent of delocalization of an individual pair of $\pi$-electrons generally is increased and suppressed, if the respective parent C=C bond belongs to a linear and cross-conjugated fragment, respectively.

## 1. Introduction

Conjugation is undoubtedly among the most important and widespread phenomena in organic chemistry. It is usually understood as an additional interaction (overlap) of $2p_z$ orbitals of atoms of unsaturated fragments (functional groups) due to formally-single interfragmental bonds. Consequently, the compounds concerned acquire specific properties, including higher stabilities as compared to appropriate reference structures (e.g. sets of the relevant isolated fragments) [1,2], shortening of interfragmental bonds vs. their standard lengths [2-4], etc. Although special types of this phenomenon often are distinguished and studied separately (e.g. the cross-conjugation [5-8]), the above-enumerated properties are more or less common to all conjugated molecules whatever the details of the structure of a particular compound. That is why the term 'the conjugation effect' also is alternatively used. It is assumed to embrace not only the preconditions (i.e. the above-specified additional interaction of $2p_z$ orbitals) but also the consequences of conjugation. These circumstances evidently stimulate theoretical investigations of various aspects of the conjugation effect, including the very nature of the latter. Numerous actual and potential applications of (poly)conjugated materials [1,5,8,9] also deserve mention here as a motive for these studies.

Acyclic unsaturated hydrocarbons (polyenes) [10] are among archetypal examples of conjugated compounds. Consequences of conjugation of double (C=C) bonds are known to be relatively weak [1,2,11,12] in this case, especially in comparison to the relevant cyclic (aromatic) analogues [1]. This especially refers to the so-called "collective" properties, i.e. those determined by the whole system of $\pi$-electrons. For example, characteristics of carbon-carbon bonds are far from being uniform over the polyene chain (cf. the so-called bond length alternation (BLA) [1,4]), although the formally-double (single) bonds are somewhat lengthened (shortened) here vs. their standard values [1-4]. The energetic effect of conjugation of C=C bonds also is rather insignificant in polyenes, viz. the conjugation energies (CEs) are small in comparison to the relevant total energies [1]. It is then no surprise that these CEs have been treated mostly in an additive manner [13-18], even more so because acyclic polyenes often served as reference systems [15-17] when studying (poly)cyclic conjugated hydrocarbons and/or their aromaticity. Nevertheless, acyclic polyenes of distinct constitutions (e.g. isomers) exhibit different properties. For example, the cross-conjugated (dendralene) chains are known to be considerably less stable vs. their linearly-conjugated counterparts [1,2,6,8,18-21]. Thus, just the acyclic conjugated hydrocarbons are under focus in the present study.

The most popular mental image of the conjugation effect follows from the predominant approach of quantum chemistry, viz. from the standard (canonical) method of molecular orbitals (MOs). This method provides us with representation of any molecular system by a set of double-occupied one-electron orbitals (i.e. canonical MOs (CMOs)) usually embracing the whole molecule concerned and thereby being completely delocalized. This especially refers to molecules consisting of similar fragments, such as polyenes. On this basis, the conjugation effect often is identified with formation of an entirely delocalized system of $\pi$-electrons (see e.g. [1,3,4]). This perspective, however, is not free from important weak points: First, the drastic delocalization of CMOs when passing from isolated C=C bonds to polyene is not compatible with the above-concluded perturbational nature of the conjugation effect in these systems. Second, delocalization of canonical MOs is due to symmetry requirements [22] and, consequently, the extent of the latter is largely independent of that of conjugation. For example, one can easily make sure that variation in relative values of resonance parameters of formally-single (C−C) bonds exerts practically no influence upon the shapes of CMOs of polyene chains in the framework of the simple Hűckel model. Moreover, such an invariance of CMOs is not in line with the common belief that the overall extent of delocalization of $\pi$-electrons grows with conjugation [It deserves recalling here that the very classification of hydrocarbons into saturated, conjugated aliphatic and aromatic often is grounded on qualitative differences in the relevant extents of delocalization (see e.g. [23])]. It is also no surprise in this context that additional characteristics of (electronic) structures are introduced and used instead of CMOs themselves in the relevant comparative studies of isomers of polyenes, e.g. the conjugation length [1,5], the topological index Z [5], graph-theoretical descriptors of branching [24], etc.

The well-known valence bond (VB) theory (see e.g. [25,26]) usually is regarded as the direct alternative to the canonical method of MOs. Applications of this theory to conjugation in general, however, are rather scarce. In our context, analysis of MO determinants of some conjugated compounds in terms of the VB theory [12] deserves mention. It has been shown that covalent bonding schemes predominate considerably over charged and radical ones. This result evidently implies a weak extent of the underlying delocalization. Again, a unified treatment of the above-specified supplementary bonding schemes via the Dewar resonance structures has been suggested recently [2]. On this basis, a general approach to conjugation has been elaborated, wherein the number of the Dewar structures serves as a measure of delocalization and thereby of conjugation.

Apart from the above-overviewed applications of the classical MO and VB theories, other approaches to conjugation also have been developed that may be jointly called the localized ones.

First of a all, approaches based on perturbative treatment of the configuration interaction (CI) deserve mention [11,21], wherein the most important configuration (the zero order many-electron wave function) coincides with the determinant built up of localized double-occupied orbitals of formally-double (C=C) bonds usually referred to as bond orbitals (BOs). In spite of the rather involved constitution of the relevant total wave function, the authors of Ref.[21] established a relation between lower stabilities of cross-conjugated polyenes (dendralenes) vs. the linear ones, on the one hand, and unfavourable conditions for the so-called indirect delocalization in the former case, on the other hand. Meanwhile, the short-range nature of delocalization in polyenes was the main conclusion of Ref.[11].

Non-canonical versions of the MO method itself [22,27] also yield localized alternatives under our interest. As opposed to the above-discussed standard (canonical) version, molecules are represented here in terms of nearly-localized (or weakly-delocalized) double-occupied one-electron orbitals (MOs), usually referred to as non-canonical MOs (NCMOs) or localized MOs (LMOs) (see e.g. [28-31] for review). The most popular way of derivation of the latter consists in transforming the relevant set of CMOs using various localization criteria [32-35] and it may be called the indirect one. Again, NCMOs (LMOs) are also obtainable directly (i.e. without any reference to CMOs) [36-46] on the basis of the Brillouin theorem [27,31,36,37] as described below. Although a few contributions has been devoted to indirect derivation of LMOs of acyclic polyenes and to their analysis (see e.g. [15,47]), the overall potential of the NCMO method for studies of conjugation is not yet sufficiently explored. The same refers even more to the potential of this method in general. Among reasons, the ambiguity of NCMOs (LMOs) is often mentioned [48]. Again, an important advantage of the NCMO method over other localized alternatives (such as the above-discussed CI-based approaches) consists in preservation of the simple and illustrative concept of the MO as a linear combination of atomic or bond orbitals. In addition, the NCMO method is in line with the classical Lewis model of localized electron pairs [49].

Difficulties with ambiguity of LMOs (NCMOs) may be largely circumvented by choice of the direct way of their derivation [46]. The point is that no "external" localization criteria are then required and, consequently, the actual delocalization of NCMOs may be expected to depend only upon the "internal" factors including the relevant extent of conjugation. Another advantage of the Brillouin theorem consists in feasibility of a powerful matrix representation both for the initial problem and for the results [40-46], the latter ultimately embracing entire classes of chemical compounds instead of individual molecules as usual. This important point deserves more detailed comments.

Among particular forms of the Brillouin theorem [27,31,36,37], there is the zero matrix requirement for the occupied-vacant off-diagonal block (submatrix) of the total one-electron Hamiltonian (Fockian) matrix ($H$) in the basis of NCMOs being sought [37-46,50]. In other words, the initial matrix $H$ should be block-diagonalized so that its final form coincides with the direct sum of two submatrices (blocks) $E_{occ}$ and $E_{vac}$, referring to subspaces of double-occupied and vacant NCMOs, respectively. The transformation matrix ($C$) obtained then contains the relevant representation of NCMOs, whereas a two-fold trace of the submatrix $E_{occ}$ (i.e. $2TrE_{occ}$) yields the respective total energy [46,50]. Given that the initial matrix $H$ also is partitioned into four submatrices (blocks) ($H_{11}$, $H_{12}$, $H_{21}$ and $H_{22}$) of appropriate dimensions (sizes), the above-mentioned block-diagonality requirement resembles a usual diagonality condition for a certain two-dimensional matrix [41], except for multi-dimensional submatrices ($E_{occ}$, $E_{vac}$, $H_{11}$, $H_{12}$, etc.) playing the role of ordinary (one-dimensional) quantities (such as eigenvalues, Coulomb and resonance parameters, etc). Thus, we actually have to do with a definite generalization of the usual diagonalization (eigenvalue) problem. It is then no surprise that solutions of the block-diagonalization problem proved to be obtainable via entire submatrices of the matrix $H$ ($H_{11}$, $H_{12}$, $H_{21}$ and $H_{22}$) without specifying either the "internal" constitutions or the dimensions of the latter [40-46,50]. This refers both to submatrices of the transformation matrix $C$ (i.e. $C_{11}$, $C_{12}$, $C_{21}$ and

$C_{22}$) and to the blocks $E_{occ}$ and $E_{vac}$ called the eigenblocks of the matrix $H$ [41]. Moreover, thanks to the well-known invariance of a matrix trace as a whole with respect to definite transformations and transpositions inside, the formalism under discussion offered new alternatives to express the total energy ($E$), in particular to relate the latter directly to shapes of NCMOs [46,50] (instead of the usual search for separate diagonal elements of the occupied eigenblock ($E_{occ}$) followed by their addition [47]). It is evident that the above-overviewed results embrace entire classes of initial Hamiltonian (Fockian) matrices and thereby of molecules.

Systems consisting of a certain number of similar weakly-interacting chemical bonds may be referred to here as an example of such a class. General solutions of the relevant block-diagonalization problem have been derived in this case in the form of power series with respect to Hamiltonian matrix blocks representing the interbond interaction [40-42,46,51]. The resulting NCMOs (LMOs) then proved to be of the bond-orbital-and-tail constitution [51], i.e. these contained the main contribution of the bonding orbital of a certain C=C bond (referred to as the parent bond of the given NCMO) and small increments (tails) extending over the nearest neighbourhood of the latter. Accordingly, total energies of these systems have been expressed in terms of the so-called delocalization coefficients of occupied NCMOs [46, 50]. Although the results of Refs. [40-42,46] originally referred to alkanes and their derivatives, $\pi$-electron systems of acyclic polyenes also belong to the class under discussion [52] provided that these hydrocarbons are modelled as sets of weakly-interacting double (C=C) bonds and the single (C–C) ones represent the interaction. Such a model, in turn, is compatible with the above-described weak conjugation effect in acyclic polyenes.

In summary, the present study is devoted to application of the above-enumerated achievements to elaboration of an alternative viewpoint of conjugation of double (C=C) bonds in acyclic polyenes in terms of nearly-localized MOs (LMOs) and thereby weakly-delocalized pairs of $\pi$-electrons. In addition, we expect to find a relation between the local delocalization pattern of these electrons and local constitution of the carbon backbone. Analysis of the relevant CEs in terms of local increments of separate electron pairs also is among our aims.

The next Section of the paper contains an overview of the principal formulae for NCMOs and total energies adapted to the case of polyenes.

## 2. Overview of expressions for NCMOs and total energies of acyclic polyenes

Let a certain polyene consist of an even number ($2N$) of uniform (carbon) atoms and contain two types of uniform bonds, namely $N$ formally-double (C=C) bonds and $N'$ formally-single (C–C) bonds, where $N$ does not coincide with $N'$. The total number of $\pi$-electrons also evidently equals to $2N$. Besides, no need arises here for specifying either the numbers $N$ and $N'$ or the overall constitution of the polyene concerned.

Further, the C–C bonds are assumed to be weak as compared to the C=C ones in accordance with the above-discussed perturbative nature of the conjugation effect in acyclic polyenes [An additional support for this assumption may be found in Ref.[52]]. On the same basis, the set of $N$ isolated C=C bonds (ethene fragments) serves as the zero order approximation (reference system) of our polyene. Meanwhile, the C–C bonds represent the interaction between the C=C bonds (perturbation). It is evident that a completely localized pair of $\pi$-electrons corresponds to each C=C bond in the zero order (reference) system. Emergence of C–C bonds as a perturbation is then expected to give rise to a certain weak delocalization of the above-mentioned pairs that is reflected in the shapes of respective NCMOs (LMOs). To find algebraic expressions for the latter, the solution of the block-diagonalization problem in the form of power series [40-42,46] will be applied. Let us now turn to an overview of these expressions.

At the Hűckel level, our polyenes are assumed to be representable by three principal parameters in the $2N$-dimensional basis of $2p_z$ atomic orbitals (AOs) of carbon atoms {$\chi$}, namely by the

Coulomb parameter ($\alpha$), as well as by resonance parameters (integrals) $\beta$ and $\gamma$ corresponding to formally-double and formally-single bonds, respectively [Parameters between AOs of more remote atoms are ignored as usual [53,54]]. Further, the parameter $\gamma$ is supposed to take a sufficiently small value so that it may be regarded as a first order term vs. $\beta$. Given that the usual equalities $\alpha = 0$ and $\beta = 1$ are accepted for convenience, a negative energy unit and thereby a positive $\gamma$ value ($\gamma > 0$) accordingly follow. Moreover, the parameter $\gamma$ is then a dimensionless quantity representing the ratio of the resonance integral of single bonds to that of double bonds.

Let us now define the bond orbitals (BOs) of formally-double C=C bonds. To this end, let the initial AOs {$\chi$} be enumerated in such a way that orbitals belonging to the same C=C bond (say, to the $I$th one) acquire the coupled numbers $i$ and $N+i$, where $i$ here and below refers to the $I$th double bond (this type of numbering is exemplified in Fig.1).

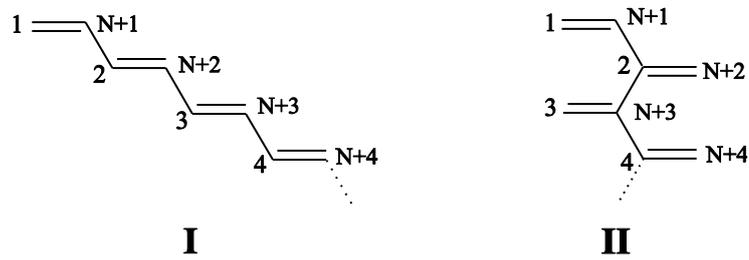

Figure 1: The linear polyene chain (I) and its cross-conjugated counterpart (II). Numberings of carbon atoms and/or of their 2p$_z$ AOs also are shown.

Let the bonding BO (BBO) of this bond $\phi_{(+)i}$ and its antibonding counterpart (ABO) $\phi_{(-)i}$ be defined as a normalized sum and difference, respectively, of the above-specified two AOs, viz.

$$\phi_{(+)i} = \frac{1}{\sqrt{2}}(\chi_i + \chi_{N+i}), \qquad \phi_{(-)i} = \frac{1}{\sqrt{2}}(\chi_i - \chi_{N+i}). \qquad (1)$$

Since passing from the basis {$\chi$} to {$\phi$} actually implies taking into account resonance parameters ($\beta = 1$) between AOs inside C=C bonds, all BBOs and all ABOs are ultimately characterized by one-electron energies correspondingly equal to 1 and –1 in our energy units, whilst the energy gap between BBOs and ABOs coincides with 2. Accordingly, resonance parameters between BOs $\phi_{(+)i}$ and $\phi_{(-)i}$ [alternatively called the intrabond ones] vanish for any $i$. Further, orbitals $\phi_{(+)i}$ and $\phi_{(-)i}$ ($i=1,2,..N$) are supposed to be double-occupied and vacant, respectively, in the zero order (reference) system. In other words, the BBOs $\phi_{(+)i}$ ($i=1,2,..N$) represent individual pairs of $\pi$-electrons of C=C bonds "before conjugation".

Let us now dwell on the analogous parameters between BOs of different C=C bonds, referred to below as interbond resonance parameters. Let us introduce the following notations for the three principal types of the latter, viz.

$$\langle \phi_{(+)i} | \hat{H} | \phi_{(+)j} \rangle = S_{ij}, \qquad \langle \phi_{(-)l} | \hat{H} | \phi_{(-)m} \rangle = Q_{lm}, \qquad \langle \phi_{(+)i} | \hat{H} | \phi_{(-)l} \rangle = R_{il}, \qquad (2)$$

where the BOs concerned are shown inside the bra- and ket-vectors. Since C=C bonds "interact" one with another via C–C ones in our model, non-zero values of parameters of Eq.(2) may be easily shown to correspond to BOs of the former connected by a C–C bond and referred to below as first-neighboring double bonds. Moreover, the absolute values of these significant parameters are uniform and equal to $\gamma/2$. As a result, parameters of Eq.(2) are first order terms with respect to $\gamma$.

Our next step consists in representing the above-defined quantities in a matrix form. Let us start with collecting all BBOs $\phi_{(+)i}$, $i=1,2,...N$ and all ABOs $\phi_{(-)i}$, $i=1,2,...N$ into two $N$-dimensional subsets of BOs correspondingly denoted by $\{\phi_{(+)}\}$ and $\{\phi_{(-)}\}$. The respective one-electron energies then accordingly compose matrices $I$ and $-I$ of the same size, where $I$ here and below stands for the unit matrix. Finally, $N$x$N$-dimensional square matrices of the first order $S$, $Q$ and $R$ may be defined that contain parameters of Eq.(2) as separate elements. Besides, diagonal elements of these matrices vanish [Note that $R_{Ii}$ coincides with the intrabond resonance parameter of the $I$th double bond]. An efficient procedure to construct matrices $S$, $Q$ and $R$ for specific systems may be found in Ref.[55].

Using the above-defined matrix representation, the total Hamiltonian matrix ($H$) of our polyene may be partitioned into four $N$x$N$-dimensional submatrices (blocks) expressible as follows [55]

$$H_{11} = I + S \qquad H_{22} = -I + Q, \qquad H_{12} = R, \qquad H_{21} = R^+, \qquad (3)$$

where the superscript + of a certain matrix here and below indicates the transposed (Hermitian-conjugate) counterpart of the latter. The blocks $H_{11}$ and $H_{22}$ contain terms of both zero ($I$) and first order ($S$ and $Q$) with respect to $\gamma$. These blocks correspond to subsets of BBOs $\{\phi_{(+)}\}$ and of ABOs $\{\phi_{(-)}\}$, respectively, and take the diagonal positions within the total matrix $H$. Meanwhile, the off-diagonal submatrices ($H_{12}$ and $H_{21}$) are first order terms vs. the former. This allowed us to express the whole subsets of double-occupied and vacant NCMOs ($\{\psi_{(+)}\}$ and $\{\psi_{(-)}\}$) as linear combinations of entire subsets $\{\phi_{(+)}\}$ and $\{\phi_{(-)}\}$, where the relevant $N$x$N$-dimensional „coefficients" consist of entire matrices $S$, $Q$ and $R$ and of their products. Formulae for individual NCMOs follow straightforwardly from these general expressions and take the form of linear combinations of both BBOs $\phi_{(+)i}$, $i=1,2,...N$ and ABOs $\phi_{(-)i}$, $i=1,2,...N$. In our context, the double-occupied NCMOs $\psi_{(+)i}$, $i=1,2,...N$ are of particular importance. Indeed, squares of the latter represent the actual distribution of the relevant pairs of $\pi$-electrons in the polyene concerned. These NCMOs take the following form

$$\psi_{(+)i} = \sum_{(+)j} \phi_{(+)j} C_{11,ji} + \sum_{(-)m} \phi_{(-)m} C_{21,mi}, \qquad (4)$$

where sums over $(+)j$ and over $(-)m$ here and below generally embrace all BBOs and all ABOs of the given system, respectively, whereas $C_{11,ji}$ and $C_{21,mi}$ stand for the relevant coefficients. As with the interbond resonance parameters of Eq.(2), the coefficients $C_{11,ji}$ and $C_{21,mi}$ coincide with elements of certain $N$x$N$-dimensional submatrices (blocks) $C_{11}$ and $C_{21}$, respectively, of the total transformation matrix $C$. The submatrix $C_{21}$ represents the so-called intersubset tails of double-occupied NCMOs, whilst $C_{11}$ reflects the consequent renormalization of the latter and has been called the renormalization matrix [46].

As already mentioned, the NCMO representation matrix (transformation matrix) $C$ has been derived in the form of power series with respect to parameters describing the interbond interaction. The same refers also to separate submatrices of the matrix $C$. In the case of polyenes, the above-introduced parameter $\gamma$ evidently underlies the expansion. Thus, submatrices $C_{11}$ and $C_{21}$ take the form of sums of increments $C_{11}^{(k)}$ and $C_{21}^{(k)}$ of increasing orders $k$ with respect to $\gamma$. Moreover, zero order double-occupied NCMOs were chosen to coincide with BBOs and, consequently, we obtain that $C_{11}^{(0)} = I$. Meanwhile, the remaining zero order increment $C_{21}^{(0)}$ vanishes. The overall form of submatrices $C_{11}$ and $C_{21}$ is then as follows

$$C_{11} = I + \sum_{k=1}^{\infty} C_{11}^{(k)}, \qquad C_{21} = \sum_{k=1}^{\infty} C_{21}^{(k)}. \qquad (5)$$

As a result, the double-occupied NCMOs $\psi_{(+)i}$ actually coincide with weakly-delocalized orbitals of double (C=C) bonds and one-to-one correspondence is preserved between these NCMOs and BBOs $\phi_{(+)i}$ in addition. That is why the $I$th C=C bond is called below the parent bond of the $I$th pair of $\pi$-electrons. Further, increments of Eq.(5) referring to $k=1,2,3...$ are representable directly via the above-defined (sub)matrices $S$, $R$ and $Q$ of the model Hamiltonian matrix of polyenes in the basis of BOs. For the sake of convenience, however, the so-called principal matrices of the PNCMO theory $G_{(k)}$, $k=1,2,3..$ [50] have been used for this purpose. The first three contributions to the intersubset block $C_{21}$ of the matrix $C$ take then the form

$$C_{21}^{(1)} = G_{(1)}, \qquad C_{21}^{(2)} = G_{(2)}, \qquad C_{21}^{(3)} = G_{(3)} + \frac{1}{2}G_{(1)}G_{(1)}^{+}G_{(1)}, \tag{6}$$

where

$$G_{(1)} = -\frac{1}{2}R, \qquad G_{(2)} = -\frac{1}{2}(SG_{(1)} - G_{(1)}Q) = \frac{1}{4}(SR - RQ),$$

$$G_{(3)} = -\frac{1}{2}(SG_{(2)} - G_{(2)}Q) - 2G_{(1)}G_{(1)}^{+}G_{(1)} \tag{7}$$

(see Eq.(2.4.2) of Ref.[50]). Besides, $G_{(k)}$, $k=1,2,3..$ were shown to be skew-symmetric (skew-Hermitian) matrices in the case of polyenes [52], i.e.

$$G_{(k)}^{+} = -G_{(k)}, \qquad G_{(k)ii} = 0, \qquad G_{(k)im} = -G_{(k)mi}, \tag{8}$$

where the superscript + indicates the transposed (Hermitian-conjugate) matrix as previously. As is seen from Eq.(7), matrices $G_{(k)}$, $k=1,2,3…$ contain factors $\gamma^{k}$ and thereby are terms of the $k$th order with respect to $\gamma$ as indicated by their subscripts ($k$). Meanwhile, the relevant contributions to the renormalization matrix $C_{11}$ take the form of sums of products of matrices $G_{(k)}$ of lower orders, i.e. of $G_{(k-1)}$, $G_{(k-2)}$, etc. Formulae for particular increments $C_{11}^{(k)}$ may be found elsewhere [46].

Let us now dwell on interpretation of elements $G_{(k)im}$ of our principal matrices $G_{(k)}$, $k=1,2,3..$. The element $G_{(1)im}$ of the first order matrix $G_{(1)}$ follows from the first relation of Eq.(7). It is proportional to the resonance parameter $R_{im}$ of Eq.(2) and inversely proportional to the energy gap between BBOs and ABOs (equal to 2 in our energy units). In this connection, $G_{(1)im}$ has been interpreted as the direct (through-space) interaction between BOs $\phi_{(+)i}$ and $\phi_{(-)m}$. As already mentioned, absolute values of resonance parameters $R_{im}$ coincide with $\gamma/2$ for pairs of first-neighboring C=C bonds I and M and vanish otherwise. Absolute values of respective significant elements $G_{(1)im}$ then accordingly equal to $\gamma/4$. It is also evident that a two-membered conjugated path [56,57] (butadiene-like fragment) corresponds to any non-zero element $G_{(1)im}$ and vice versa. [This paths is abbreviated below by CP(2)]. Analogously, the element $G_{(2)im}$ of the second order matrix $G_{(2)}$ is defined by the second relation of Eq.(7). Analysis of this formula [55] shows that the element $G_{(2)im}$ represents the indirect (through-bond) interaction of BOs $\phi_{(+)i}$ and $\phi_{(-)m}$ by means of other basis orbitals of the given system playing the role of mediators. To be an efficient mediator, however, a certain BO (i.e. either a BBO $\phi_{(+)j}, j \neq i$ or an ABO $\phi_{(-)l}, l \neq m$) should overlap with both $\phi_{(+)i}$ and $\phi_{(-)m}$ and thereby it should belong to a common first neighbour of the C=C bonds concerned (i.e. I and M). Given that the latter really possess a common first neighbour, these will be called second-neighboring C=C bonds. Thus, pertinence of BOs $\phi_{(+)i}$ and $\phi_{(-)m}$ to second-neighboring C=C bonds is a necessary condition for the element $G_{(2)im}$ to take a non-zero value. Apart from this necessary condition, however, the mutual arrangement of the $I$th and $M$th C=C bonds with respect to the mediating one(s) (say, the $L$th C=C bond) starts to play an equally important role. Indeed, a non-zero value of the element $G_{(2)im}$ was shown to be ensured [55] only if the $I$th, $L$th and $M$th C=C bonds are linearly conjugated (but not cross-conjugated). In other words, coincidence of the fragment I–L–M with a three-membered conjugated path [abbreviated below as CP(3)] is required here.

Let us now turn to specific characteristics of NCMOs of Eq.(4) introduced previously [41,42,46] to represent the extents of their delocalization. Let $d_{(+)i,(-)m}$ stand for the partial (intersubset) delocalization coefficient of the NCMO $\psi_{(+)i}$ over the ABO $\phi_{(-)m}$ defined as square of the relevant coefficient $C_{21,mi}$ of Eq.(4). As with the latter, the partial delocalization coefficient $d_{(+)i,(-)m}$ also takes the form of an analogous power series, i.e.

$$d_{(+)i,(-)m} = |C_{21,mi}|^2 = \sum_{k=2}^{\infty} d^{(k)}_{(+)i,(-)m}, \qquad (9)$$

where $d^{(k)}_{(+)i,(-)m}$ stands for the increment of the $k$th order with respect to $\gamma$. For $k=2,3$ and 4, we correspondingly obtain [46]

$$d^{(2)}_{(+)i,(-)m} = (G_{(1)im})^2 > 0, \qquad d^{(3)}_{(+)i,(-)m} = 2G_{(1)im}G_{(2)im},$$
$$d^{(4)}_{(+)i,(-)m} = 2G_{(1)im}G_{(3)im} + G_{(1)im}(G_{(1)}G_{(1)}^+G_{(1)})_{im} + (G_{(2)im})^2, \qquad (10)$$

where $G_{(k)im}$ are individual elements of matrices $G_{(k)}$ resulting from Eq.(7). Accordingly, the notation $(G_{(1)}G_{(1)}^+G_{(1)})_{im}$ here and below stands for the relevant element of the matrix product $G_{(1)}G_{(1)}^+G_{(1)}$ [Note that the first order increment $d^{(1)}_{(+)i,(-)m}$ vanishes when building up the square of the coefficient $C_{21,mi}$]. The *a priori* positive sign of the second order term $d^{(2)}_{(+)i,(-)m}$ also deserves emphasizing. Meanwhile, the signs of terms of higher orders $d^{(k)}_{(+)i,(-)m}$, $k=3,4,..$ cannot be established *a priori*.

Let us now dwell on some general properties of coefficients $d^{(k)}_{(+)i,(-)m}$. The skew-symmetric nature of matrices $G_{(k)}$ (see Eq.(8)) evidently is extendable to the matrix product $G_{(1)}G_{(1)}^+G_{(1)}$. Consequently, the equality $C^{(k)}_{21,ii} = 0$ follows from Eq.(6) for any $k$. Thus, a double-occupied NCMO $\psi_{(+)i}$ of Eq.(4) contains no increment of the ABO of the same C=C bond ($\phi_{(-)i}$). In other terms, the "heads" [51] of NCMOs of polyenes are of zero polarities. As is seen from Eq.(10), the intrabond delocalization coefficients $d^{(k)}_{(+)i,(-)i}$ also vanish for any $k$. Finally, separate increments of the right-hand sides of relations of Eq.(10) are no exceptions in this respect. Further, the skew-symmetric nature of matrices $G_{(k)}$ yields the equalities

$$d^{(k)}_{(+)i,(-)m} = d^{(k)}_{(+)m,(-)i} \qquad (11)$$

for any $k$ whatever the actual mutual arrangement of the $I$th and $M$th C=C bonds. Hence, the extent of delocalization of the NCMO $\psi_{(+)i}$ over the ABO of the $M$th bond ($\phi_{(-)m}$) coincides with that of the $M$th NCMO ($\psi_{(+)m}$) over the ABO $\phi_{(-)i}$. This result is referred to below as the symmetry property of partial delocalization coefficients. As with the above-discussed zero polarity of the "heads" of NCMOs, the symmetry property is extendable to separate increments of relations of Eq.(10). Besides, the result of Eq.(11) was shown to give rise to zero interbond charge transfer in polyenes [50] and causes little surprise.

Let us now define the total (intersubset) delocalization coefficient of the same NCMO ($\psi_{(+)i}$) over all ABOs as follows

$$D_{(+)i} = \sum_{(-)m} d_{(+)i,(-)m}. \qquad (12)$$

Substituting the right relation of Eq.(9) into Eq.(12) yields an expression for $D_{(+)i}$ in the form of power series, i.e.

$$D_{(+)i} = \sum_{k=2}^{\infty} D^{(k)}_{(+)i}, \qquad (13)$$

where

$$D^{(k)}_{(+)i} = \sum_{(-)m} d^{(k)}_{(+)i,(-)m} \qquad (14)$$

and $D^{(2)}_{(+)i} > 0$. Besides, total (intersubset) delocalization coefficients of NCMOs ($D_{(+)i}$) were shown to determine the actual bond orders inside the formally-double (C=C) bonds in polyenes [50]. In particular, any positive (negative) term of the expansion of Eq.(13) is accompanied by a negative (positive) increment to the "internal" order of the $I$th C=C bond and thereby contributes to weakening (strengthening) of the latter as compared to an isolated C=C bond. The *a priori* positive sign of the second order term $D^{(2)}_{(+)i}$ then ensures the ultimate weakening of the $I$th C=C bond due to conjugation and this outcome causes little surprise. An important point here also is that intrasubset tails of NCMOs (i.e. the coefficients $C_{11,ji}$ of Eq.(4)) do not participate in the formation of the above-discussed "internal" characteristics of formally-double (C=C) bonds. This fact is among reasons why intrasubset tails are not included into definitions of partial and total delocalization coefficients of NCMOs [Additional arguments for this option are given below]. Finally, the overall extent of delocalization of double-occupied NCMOs $\psi_{(+)i}$, $i=1,2\dots N$ may be represented by the following quantity

$$D_{(+)} = \sum_{(+)i} D_{(+)i} \tag{15}$$

that has been called the complete delocalization coefficient of double-occupied NCMOs. It is evident that $D_{(+)}$ also is representable as a sum over $k$ of increments $D^{(k)}_{(+)}$ expressible as follows

$$D^{(k)}_{(+)} = \sum_{(+)i} \sum_{(-)m} d^{(k)}_{(+)i,(-)m} . \tag{16}$$

Let us now turn to the relevant total energy ($E$). As already mentioned, the energy $E$ is defined as a two-fold trace of the occupied eigenblock ($\boldsymbol{E}_{occ}$) of the Hamiltonian matrix ($\boldsymbol{H}$) in the method applied. As with the above-discussed submatrices $\boldsymbol{C}_{11}$ and $\boldsymbol{C}_{21}$ of Eq.(5), the eigenblocks $\boldsymbol{E}_{occ}$ and $\boldsymbol{E}_{vac}$ also are expressible in the form of an analogous power series [46,50,58], i.e. as sums over $k$ of increments $\boldsymbol{E}^{(k)}_{occ}$ and $\boldsymbol{E}^{(k)}_{vac}$, the latter consisting of products of our principal matrices $\boldsymbol{R}$, $\boldsymbol{G}_{(1)}$, $\boldsymbol{G}_{(2)}$, etc. The same then consequently refers to the relevant total energy ($E$), the $k$th order member of the expansion for which ($E_{(k)}$) is defined as follows

$$E_{(k)} = 2Tr\boldsymbol{E}^{(k)}_{occ} , \tag{17}$$

where the right-hand side of Eq.(17) contains a trace (denoted here and below by $Tr$) of the $k$th order member of the expansion for the eigenblock $\boldsymbol{E}_{occ}$. Expressions for energy increments $E_{(k)}$ may be found elsewhere [50,55,59]. Thus, we confine ourselves here to a brief overview of the principal results.

The zero order increment ($E_{(0)}$) coincides with the double-sum of one-electron energies of BBOs $\phi_{(+)i}$, $i=1,2,3\dots N$ and thereby with the total energy of $N$ isolated double bonds equal to $2N$ in our energy units. Meanwhile, the first order contribution of the same series ($E_{(1)}$) was shown to vanish [46,50,59]. Furthermore, cyclic transpositions of matrix factors (i.e. of $\boldsymbol{R}$, $\boldsymbol{G}_{(1)}$, $\boldsymbol{G}_{(2)}$, etc) inside the trace signs of formulae for the increments $E_{(k)}$ of higher orders ($k=2, 3,..$) allowed the latter to be expressed via increments of the same order to the above-defined (intersubset) delocalization coefficients of NCMOs, viz.

$$E_{(k)} = \frac{4}{k-1} \sum_{(+)i} \sum_{(-)m} d^{(k)}_{(+)i,(-)m} = \frac{4}{k-1} \sum_{(+)i} D^{(k)}_{(+)i} = \frac{4}{k-1} D^{(k)}_{(+)} \equiv \frac{4}{k-1} Tr\boldsymbol{D}^{(k)}_{(+)} , \tag{18}$$

[see Appendix C and Eq.(43) of Ref.[46] under an assumption that all the energy intervals $\varepsilon_{(+)i} + \varepsilon_{(-)l}$ are uniform and equal to 2]. Notations $\boldsymbol{D}^{(k)}_{(+)}$ of the right relation of Eq.(18) stand for the so-called intersubset delocalization matrices, diagonal elements of which ($D^{(k)}_{(+)ii}$) coincide with total delocalization coefficients of the $k$th order ($D^{(k)}_{(+)i}$). It is evident that the conjugation energy (CE) of a certain polyene ($\Delta E$) may be defined as a difference between the total energy ($E$) and the respective zero order increment ($E_{(0)}$). The series for $\Delta E$ then starts with the term of the second order, viz.

$$\Delta E = \sum_{k=2}^{\infty} E_{(k)} = E_{(2)} + E_{(3)} + E_{(4)} + ..., \qquad (19)$$

where $E_{(k)}$ are shown in Eq.(18). Hence, the CE of polyene is actually determined by the overall extent of intersubset delocalization of double-occupied NCMOs and thereby it is interpretable as the total (intersubset) delocalization energy of all pairs of $\pi$-electrons of the given polyene. It deserves a separate mention that the intrasubset delocalization is not contained in Eqs.(18) and (19) as it was the case with "internal" bond orders of formally-double (C=C) bonds. This result evidently causes little surprise if we recall the well-known zero energetic effect of interaction of double-occupied orbitals (see e.g. [60]). The less important nature of intrasubset tails of NCMOs is supported also by expressibility of renormalization matrices $C_{11}^{(k)}$ via the intersubset delocalization matrices $D_{(+)}^{(k)}$ [46].

Additivity of the CE of polyene with respect to contributions of individual pairs of $\pi$-electrons also is among outcomes of Eqs.(18) and (19). Indeed, sums over $k$ and over $(+)i$ may be interchanged in these relations and we then obtain that

$$\Delta E = \sum_{I} \Delta E_{I} , \qquad (20)$$

where the contribution of the $I$th pair of electrons ($\Delta E_{I}$) is related to the overall extent of (intersubset) delocalization of the respective single NCMO $\psi_{(+)i}$, viz.

$$\Delta E_{I} = \sum_{k=2}^{\infty} \Delta E_{I(k)} = \sum_{k=2}^{\infty} \frac{4}{k-1} D_{(+)i}^{(k)} = 4 D_{(+)i}^{(2)} + 2 D_{(+)i}^{(3)} + \frac{4}{3} D_{(+)i}^{(4)} + ..., \qquad (21)$$

Thus, the more delocalized the $I$th pair of $\pi$-electrons becomes, the higher its contribution to the total CE gets and vice versa. In other words, Eq.(21) represents a direct relation between the extent of delocalization of the $I$th pair of $\pi$-electrons and its contribution to the total CE. It deserves emphasizing here that separate increments $\Delta E_{I}$ generally do not coincide with respective diagonal elements of the occupied eigenblock $E_{occ}$ and thereby with one-electron energies of double-occupied NCMOs. Besides, the usual way of derivation of total energies consists in finding the one-electron energies of individual NCMOs (LMOs) followed by summing them up (see e.g. [47]) in the NCMO method too. Moreover, interrelations are often anticipated between the above-specified energies and the shapes of NCMOs [47,61] (although no explicit forms of these relations are known). In this context, Eqs. (19)-(21) offer a direct relation between the CE of polyene and the shapes of its NCMOs with no reference to one-electron energies of the latter.

Another perspective on the results of Eqs.(19)-(21) follows after recalling that any double-occupied NCMO $\psi_{(+)i}$ possesses a parent double (C=C) bond (viz. the $I$th one). The increment $\Delta E_{I}$ of Eq.(21) may be then accordingly interpreted as the contribution of the $I$th C=C bond to the total CE. On the same basis, Eq.(20) points to additivity of the CE with respect to increments of these bonds. Hence, the additive schemes of Refs. [15-18] for the CEs of polyenes acquire an additional support. One can also conclude that the lower the actual order inside the $I$th C=C bond becomes, the more significant its participation in the formation of the CE gets and vice versa (the above-discussed relation of $D_{(+)i}$ to "internal" bond orders should be recalled here).

Both partial and total delocalization coefficients of NCMOs of polyenes along with their CEs are likely to depend upon specific constitutions of the given molecule. This dependence is contained implicitly in the expressions for $d_{(+)i,(-)m}^{(k)}$ of Eq.(10) and it will be studied in the next Section in a detail.

## 3. Analysis of the relation between the delocalization pattern and constitution of the carbon backbone

Let us start with the second order terms ($k=2$). The positive increment $d^{(2)}_{(+)i,(-)m}$ of Eq.(10) is determined by square of the element $G_{(1)im}$. As already mentioned, the absolute value of this element coincides with $\gamma/4$ for any pair of first-neighboring C=C bonds and/or for any CP(2) and vanishes otherwise. An analogous property refers also to the increments $d^{(2)}_{(+)i,(-)m}$. We then obtain that

$$d^{(+2f)}_{(+)i,(-)m} = d^{(+2f)}_{(+)m,(-)i} = \frac{\gamma^2}{16} > 0, \tag{22}$$

where the additional superscripts + and $f$ correspondingly indicate the positive sign of the increment concerned and its relevance to first-neighboring C=C bonds. Second order contributions to the total delocalization coefficient of the NCMO $\psi_{(+)i}$ and to the overall CE of the given polyene then follow from Eqs.(14) and (18) and take the form

$$D^{(2)}_{(+)i} = \frac{\gamma^2}{16} n_I > 0, \qquad E_{(2)} = \frac{\gamma^2 N'}{2} > 0, \tag{23}$$

where $n_I$ is the number of first neighbours of the $I$th C=C bond. As is seen from the first relation of Eq.(23), the positive increment of the second order to delocalization is entirely local in nature and depends only upon constitution of the nearest (first) neighbourhood of the given C=C bond. More precisely, the increment $D^{(2)}_{(+)i}$ represents all C−C bonds attached to the $I$th C=C bond and thereby all CP(2)s embracing the latter. Meanwhile, the main outcome of the second relation consists in coincidence of second order energies for isomers [55] and proportionalities of these energies to total numbers of both C−C bonds and CP(2)s present there.

As opposed to $d^{(2)}_{(+)i,(-)m}$, the definition of $d^{(3)}_{(+)i,(-)m}$ of Eq.(10) contains also the element $G_{(2)im}$ of the second order matrix $G_{(2)}$ expressed by the second relation of Eq.(7). As discussed already, a necessary condition for the element $G_{(2)im}$ to take a non-zero value consists in pertinence of BOs concerned ($\phi_{(+)i}$ and $\phi_{(-)m}$) to second-neighboring C=C bonds. Since no pairs of C=C bonds are present in acyclic polyenes that are both first and second neighbours simultaneously, the increments $d^{(3)}_{(+)i,(-)m}$ vanish. The same then refers also to third order contributions ($D^{(3)}_{(+)i}$) to total delocalization coefficients of NCMOs, as well as to respective increments to the total CEs of polyenes [55].

Let us now turn to increments of the fourth order to partial delocalization coefficients, viz. to $d^{(4)}_{(+)i,(-)m}$ defined by the last relation of Eq.(10). The third order element $G_{(3)im}$ of this definition may be easily eliminated by substituting the last relation of Eq.(7). The result is then as follows

$$d^{(4)}_{(+)i,(-)m} = -G_{(1)im}(SG_{(2)} - G_{(2)}Q)_{im} - 3G_{(1)im}(G_{(1)}G^+_{(1)}G_{(1)})_{im} + (G_{(2)im})^2. \tag{24}$$

Let us assume first that the orbitals $\phi_{(+)i}$ and $\phi_{(-)m}$ belong to first-neighboring C=C bonds I and M. The first order element $G_{(1)im}$ then takes a non-zero value, whereas the second order one ($G_{(2)im}$) vanishes as discussed above. Thus, the last increment of Eq.(24) also vanishes in this case and we obtain

$$d^{(4f)}_{(+)i,(-)m} = d^{(+4f)}_{(+)i,(-)m} + d^{(-4f)}_{(+)i,(-)m}, \tag{25}$$

where

$$d^{(+4f)}_{(+)i,(-)m} = -G_{(1)im}(SG_{(2)} - G_{(2)}Q)_{im}, \tag{26}$$

$$d^{(-4f)}_{(+)i,(-)m} = -3G_{(1)im}(G_{(1)}G^+_{(1)}G_{(1)})_{im}. \tag{27}$$

The superscript $f$ serves to indicate the relevance of the increment concerned to first-neighboring C=C bonds, whereas the superscripts + and − specify the signs of subcomponents of Eq.(25)

established below. Given that I and M are second neighbours, a non-zero value of the element $G_{(2)im}$ is allowed but not of $G_{(1)im}$. As a result, only the last term of Eq.(24) remains and we obtain

$$d_{(+)i,(-)m}^{(+4s)} = (G_{(2)im})^2 > 0, \qquad (28)$$

where the superscript $s$ now accordingly refers to second-neighbouring C=C bonds. Finally, $d_{(+)i,(-)m}^{(4)}$ vanishes for more distant C=C bonds as both $G_{(1)im}$ and $G_{(2)im}$ take zero values in this case. Besides, an analogous partition of the fourth order energy ($E_{(4)}$) also is easily obtainable by employment of the first relation of Eq.(18). The relevant components will be accordingly denoted by $E_{(4)}^{(+f)}$, $E_{(4)}^{(-f)}$ and $E_{(4)}^{(+s)}$. Let us now consider the increments of Eqs.(25)-(28) separately.

Let us start with components of Eqs.(25)-(27), where I and M are first-neighboring C=C bonds. The component $d_{(+)i,(-)m}^{(+4f)}$ is representable as follows

$$d_{(+)i,(-)m}^{(+4f)} = -G_{(1)im} \sum_{(+)j \neq i,m} S_{ij} G_{(2)jm} + G_{(1)im} \sum_{(-)l \neq i,m} G_{(2)il} Q_{lm}, \qquad (29)$$

where sums over (+)$j$ and over (−)$l$ correspondinly embrace BBOs and ABOs of the given system except for orbitals of the $I$th and $M$th C=C bonds (viz. $\phi_{(+)i}$, $\phi_{(+)m}$ and $\phi_{(-)i}$, $\phi_{(-)m}$, respectively). The exception is due to the above-discussed zero diagonal (intrabond) elements of matrices $S$, $Q$ and $G_{(2)}$. Moreover, the sum over (+)$j$ actually embraces BBOs of only first neighbours of the $I$th C=C bond (except for the $M$th one) because of zero values of off-diagonal elements of the matrix $S$ for more distant C=C bonds. Similarly, the second sum of Eq.(29) concerns ABOs of first neighbours (L) of the $M$th bond, where L does not coincide with I. Again, the elements $G_{(2)jm}$ and $G_{(2)il}$ of Eq.(29) do not vanish provided that BOs $\phi_{(+)j}$ and $\phi_{(-)m}$, as well as $\phi_{(+)i}$ and $\phi_{(-)l}$ belong to second-neighboring C=C bonds that are terminals of at least a single CP(3) in addition. This implies that the $J$th first neighbour of the $I$th C=C bond ($J \neq M$) and/or the $L$th first neighbour of the $M$th C=C bond ($L \neq I$) contribute to the component $d_{(+)i,(-)m}^{(+4f)}$, if the triplets of C=C bonds J, I and M and/or I, M and L are linearly conjugated, i.e. if fragments J–I–M and/or I–M–L coincide with CP(3)s. In other words, participation of a $J$th first neighbour of the $I$th C=C bond ($J \neq M$) in a CP(3) is a necessary condition for the component $d_{(+)i,(-)m}^{(+4f)}$ to take a non-zero value. The same evidently refers also to the $L$th first neighbour of the $M$th C=C bond, where $L \neq I$. For example, the bonds both C$_1$=C$_5$ and C$_4$=C$_8$ of the linear octatetraene I($N$=4) contribute to $d_{(+)2,(-)3}^{(+4f)}$ due to their participation in CP(3)s C$_1$=C$_5$–C$_2$=C$_6$–C$_3$=C$_7$ and C$_2$=C$_6$–C$_3$=C$_7$–C$_4$=C$_8$, respectively (Fig. 1). Additivity of the overall component $d_{(+)i,(-)m}^{(+4f)}$ with respect to increments of several CP(3)s (if any) also easily follows from Eq.(29). Finally, similarity of all CP(3)s in polyenes automatically ensures uniform values of contributions of individual representatives of the latter that may be easily shown to coincide with $\gamma^4/64$. This evidently implies the positive sign of the component concerned whatever the actual number of CP(3)s and this fact is reflected by the superscript + of the increment $d_{(+)i,(-)m}^{(+4f)}$. Accordingly, the energetic effect of this component is stabilizing, i.e. $E_{(4)}^{(+f)} > 0$.

Let us now consider the second component of Eq.(25), namely $d_{(+)i,(-)m}^{(-4f)}$ defined by Eq.(27), where $\phi_{(+)i}$ and $\phi_{(-)m}$ also belong to first-neighboring C=C bonds I and M. The element $(G_{(1)}G_{(1)}^+G_{(1)})_{im}$ of Eq.(27) may be alternatively represented as follows

$$(G_{(1)}G_{(1)}^+G_{(1)})_{im} = \sum_{(+)j} (G_{(1)}G_{(1)}^+)_{ij} G_{(1)jm} \equiv \sum_{(-)l} G_{(1)il} (G_{(1)}^+G_{(1)})_{lm}, \qquad (30)$$

where sums over (+)$j$ and over (−)$l$ correspondingly embrace BBOs and ABOs as previously. With this in mind, the relation of Eq.(27) itself may be reformulated so that it resembles Eq.(29), viz.

$$d^{(-4f)}_{(+)i,(-)m} = -\frac{3}{2}G_{(1)im}\sum_{(+)j\neq m}(G_{(1)}G^+_{(1)})_{ij}G_{(1)jm} - \frac{3}{2}G_{(1)im}\sum_{(-)l\neq i}G_{(1)il}(G^+_{(1)}G_{(1)})_{lm}, \qquad (31)$$

where elements of matrices $G_{(1)}G^+_{(1)}$, $G^+_{(1)}G_{(1)}$ and $G_{(1)}$ now stand instead of those of the former matrices $S$, $Q$ and $G_{(2)}$. In this connection, an important difference of Eq.(31) from Eq.(29) consists in the absence of restrictions $(+)j \neq (+)i$ and $(-)l \neq (-)m$ in the former case. The reason is that products $G_{(1)}G^+_{(1)}$ and $G^+_{(1)}G_{(1)}$ are symmetric matrices and thereby contain non-zero diagonal elements $(G_{(1)}G^+_{(1)})_{ii}$ and $(G^+_{(1)}G_{(1)})_{mm}$ [55,62]. The increments of Eq.(31) embracing the latter may be then distinguished from those referring to $(+)j \neq (+)i$ and $(-)l \neq (-)m$ by partitioning the overall component $d^{(-4f)}_{(+)i,(-)m}$ into two subcomponents denoted by additional superscripts $a$ and $b$, viz.

$$d^{(-4f)}_{(+)i,(-)m} = d^{(-4f)a}_{(+)i,(-)m} + d^{(-4f)b}_{(+)i,(-)m}, \qquad (32)$$

where

$$d^{(-4f)a}_{(+)i,(-)m} = -\frac{3}{2}(G_{(1)im})^2\left[(G_{(1)}G^+_{(1)})_{ii} + (G^+_{(1)}G_{(1)})_{mm}\right], \qquad (33)$$

$$d^{(-4f)b}_{(+)i,(-)m} = -\frac{3}{2}G_{(1)im}\sum_{(+)j\neq i,m}(G_{(1)}G^+_{(1)})_{ij}G_{(1)jm} - \frac{3}{2}G_{(1)im}\sum_{(-)l\neq i,m}G_{(1)il}(G^+_{(1)}G_{(1)})_{lm}. \qquad (34)$$

Let us now discuss the subcomponents of Eqs.(33) and (34) separately.

Let us start with $d^{(-4f)a}_{(+)i,(-)m}$ of Eq.(33). Diagonal elements of matrix products $G_{(1)}G^+_{(1)}$ and $G^+_{(1)}G_{(1)}$ were shown to be determined by first neighbourhoods of the C=C bonds concerned in the case of polyenes [55]. More precisely, the elements $(G_{(1)}G^+_{(1)})_{ii}$ and $(G^+_{(1)}G_{(1)})_{mm}$ are proportional to total numbers of first neighbours $n_I$ and $n_M$ of the $I$th and $M$th C=C bond, respectively, i.e.

$$(G_{(1)}G^+_{(1)})_{ii} = \frac{\gamma}{16}n_I, \qquad (G^+_{(1)}G_{(1)})_{mm} = \frac{\gamma}{16}n_M. \qquad (35)$$

It is evident that the $M$th C=C bond and the $I$th one necessarily are included when counting the first neighbours of the $I$th and $M$th C=C bonds, respectively. Consequently, a definite "own" increment of our principal pair of C=C bonds I and M arises in the relation of Eq.(33) that is twice as large as that of any other first neighbour of either $I$th and $M$th bond. Let $n_{I(\neq M)}$ and $n_{M(\neq I)}$ stand for numbers of other first neighbours of the $I$th and $M$th C=C bond, where $n_{I(\neq M)} = n_I - 1$ and $n_{M(\neq I)} = n_M - 1$. If we recall now that elements $G_{(1)im}$ take uniform absolute values equal to $\gamma/4$ for any pair of first neighbours I and M, from Eq.(33) we obtain

$$d^{(-4f)a}_{(+)i,(-)m} = -\frac{3\gamma^4}{512}(2 + n_{I(\neq M)} + n_{M(\neq I)}). \qquad (36)$$

Let us now turn to the subcomponent $d^{(-4f)b}_{(+)i,(-)m}$ of Eq.(34) containing the off-diagonal elements $(G_{(1)}G^+_{(1)})_{ij}$ and $(G^+_{(1)}G_{(1)})_{lm}$ of matrix products $G_{(1)}G^+_{(1)}$ and $G^+_{(1)}G_{(1)}$. These elements were shown to take non-zero values for any pair of second-neighboring C=C bonds (irrespective of their type of conjugation) [55]. It is evident that any first neighbour (L) of the $M$th C=C bond ($L \neq I$) automatically is a second neighbour of the $I$th C=C bond. The same refers also to the $J$th first neighbour of the $I$th C=C bond ($J \neq M$). Moreover, contributions of these additional first neighbours of both C=C bonds concerned (I and M) take uniform values. The final expression for the subcomponent $d^{(-4f)b}_{(+)i,(-)m}$ is then as follows

$$d^{(-4f)b}_{(+)i,(-)m} = -\frac{3\gamma^4}{512}(n_{I(\neq M)} + n_{M(\neq I)}). \qquad (37)$$

After summing up the subcomponents of Eqs.(36) and (37) in accordance with Eq.(32), we obtain

$$d^{(-4f)}_{(+)i,(-)m} = d^{(-4f)}_{(+)m,(-)i} = -\frac{3\gamma^4}{256}(1 + n_{I(\neq M)} + n_{M(\neq I)}). \tag{38}$$

It is seen that $d^{(-4f)}_{(+)i,(-)m}$ always is a negative quantity as indicated by its superscript $-$. This, in turn, implies a negative (destabilizing) energetic effect of this component, i.e $E^{(-f)}_{(4)} < 0$. The symmetry property of the left-hand side of Eq.(38) is due to the skew-symmetric nature of matrices both $\mathbf{G}_{(1)}$ and $\mathbf{G}_{(1)}\mathbf{G}^+_{(1)}\mathbf{G}_{(1)}$ contained in the relevant definition of Eq.(27). More importantly, the increment $d^{(-4f)}_{(+)i,(-)m}$ of Eq.(38) takes a non-zero value for any pair of first-neighboring C=C bonds (in contrast to the above-considered component $d^{(+4f)}_{(+)i,(-)m}$, where participation of the $I$th and $M$th C=C bonds in a certain CP(3) was additionally required). Hence, the component $d^{(-4f)}_{(+)i,(-)m}$ may be interpreted as representing a certain universal repulsion of electron pairs of first-neighboring C=C bonds that contributes to destabilization of the system in a certain analogy with the ideas underlying the VSEPR model [63,64]. Moreover, the component $d^{(-4f)}_{(+)i,(-)m}$ grows with increasing number of other first neighbours of the bonds concerned (i.e. of the sum $n_{I(\neq M)} + n_{M(\neq I)}$).

The overall fourth order increment ($d^{(4f)}_{(+)i,(-)m}$) to the partial delocalization coefficient of the NCMO $\psi_{(+)i}$ of the $I$th bond over the ABO of the first-neighboring ($M$th) bond ($\phi_{(-)m}$) follows after summing up the components $d^{(+4f)}_{(+)i,(-)m}$ and $d^{(-4f)}_{(+)i,(-)m}$ in accordance with Eq.(25). Thus, $d^{(4f)}_{(+)i,(-)m}$ generally consists of two components of opposite signs and of comparable absolute values. Moreover, the above-established dependence of both components upon constitutions of first neighbourhoods of C=C bonds concerned (I and M) ensures an analogous dependence of the overall fourth order increment $d^{(4f)}_{(+)i,(-)m}$ and thereby a semi-local nature of the latter. If we recall finally, that all pairs of first-neighboring C=C bonds are characterized by uniform positive second order contributions $d^{(+2f)}_{(+)i,(-)m}$ to delocalization (see Eq.(22)), the above-discussed terms $d^{(4f)}_{(+)i,(-)m}$ actually give rise to certain alterations of the former depending on the actual constitutions of the relevant first neighbourhoods.

By contrast, pairs of second-neighboring C=C bonds (I and M) are characterized only by increments of the fourth order to delocalization defined by Eq.(28). The *a priori* positive signs of all increments $d^{(+4s)}_{(+)i,(-)m}$ are evident. As discussed already, the second order elements $G_{(2)im}$ ($i \neq m$) contained within Eq.(28) take non-zero values if the $I$th and $M$th C=C bonds coincide with terminals of a CP(3), e.g. of I–L–M. For example, this condition is met for bonds $C_1=C_5$ and $C_3=C_7$ of the linear octatetraene I(N=4). Moreover, the absolute values of elements $G_{(2)im}$ are uniform under the above-specified conditions and equal to $\gamma^2/8$. The increment $d^{(+4s)}_{(+)i,(-)m}$ then accordingly coincides with $\gamma^4/64$. Due to the symmetry property of partial delocalization coefficients and their components, the same value refers also to $d^{(+4s)}_{(+)m,(-)i}$. Finally, the stabilizing nature of the respective energetic effect ($E^{(+s)}_{(4)}$) deserves mention. Besides, the sum $E^{(+f)}_{(4)} + E^{(+s)}_{(4)}$ may be easily shown to coincide with the positive component $E^{(+)}_{(4)}$ of the fourth order energy $E_{(4)}$ revealed earlier [55,62]. Similarly, $E^{(-f)}_{(4)}$ is nothing more than the relevant negative component $E^{(-)}_{(4)}$ of Refs.[55,62].

Let us now turn to the increment of the fourth order ($D^{(4)}_{(+)i}$) to the total delocalization coefficient ($D_{(+)i}$) of the NCMO $\psi_{(+)i}$. This increment follows from Eq.(14), where the sum over ($-$

$)m$ embraces the ABOs of only first and second neighbours of the $I$th C=C bond (but not the more distant ones). It is also evident that contributions $d_{(+)i,(-)m}^{(+4f)}$, $d_{(+)i,(-)m}^{(+4s)}$ and $d_{(+)i,(-)m}^{(-4f)}$ may be summed up separately and independently. We then obtain

$$D_{(+)i}^{(4)} = D_{(+)i}^{(+4)} + D_{(+)i}^{(-4)}, \qquad (39)$$

where the positive and negative components of the correction $D_{(+)i}^{(4)}$ are denoted by $D_{(+)i}^{(+4)}$ and $D_{(+)i}^{(-4)}$, respectively. The former is related to participation of the $I$th C=C bond in CP(3)s and contains sums over first and over second neighbours of increments $d_{(+)i,(-)m}^{(+4f)}$ and $d_{(+)i,(-)m}^{(+4s)}$, respectively, viz.

$$D_{(+)i}^{(+4)} = \sum_{(-)m(f)} d_{(+)i,(-)m}^{(+4f)} + \sum_{(-)m(s)} d_{(+)i,(-)m}^{(+4s)}. \qquad (40)$$

It is evident that participation of a certain second neighbour (say, of the $L$th one) in the formation of $D_{(+)i}^{(+4)}$ depends only on whether the first neighbourhood of the $I$th bond ensures at least a single CP(3) between the $I$th and $L$th bonds. Hence, the component $D_{(+)i}^{(+4)}$ actually is determined by constitutions of only the first and second neighbourhoods of the parent bond ($I$). Meanwhile, the negative component $D_{(+)i}^{(-4)}$ is expressible as follows

$$D_{(+)i}^{(-4)} = \sum_{(-)m(f)} d_{(+)i,(-)m}^{(-4f)} \qquad (41)$$

and represents the overall repulsion undergone by the $I$th pair of $\pi$-electrons due to the presence of its first neighbours. After employment of Eq.(38) we obtain that

$$D_{(+)i}^{(-4)} = -\frac{3\gamma^4}{256}\left[n_I^2 + \sum_M n_{M(\neq I)}\right], \qquad (42)$$

where the sum over M now embraces all first neighbours of the $I$th C=C bond [Note that $n_I$ and $n_I \times (n_I - 1)$ result after summing up the first and the second term of Eq.(38), respectively]. It is evident that the sum of the right-hand side of Eq.(42) yields the total number of second neighbours of the $I$th C=C bond. Hence, the negative component $D_{(+)i}^{(-4)}$ also is determined by constitution only of the first and of the second neighbourhoods of the C=C bond concerned.

In summary, the total delocalization coefficient ($D_{(+)i}$) of the NCMO $\psi_{(+)i}$ is expressible as follows

$$D_{(+)i} = D_{(+)i}^{(+2)} + D_{(+)i}^{(+4)} + D_{(+)i}^{(-4)}, \qquad (43)$$

where the first term of the right-hand side coincides with the positive second order correction $D_{(+)i}^{(2)}$ of Eq.(23). As discussed already, this increment is associated with all C−C bonds attached to the $I$th C=C bond and thereby with all CP(2)s embracing the latter. The local origin of corrections $D_{(+)i}^{(2)}$ also deserves recalling here. Finally, the latter always take non-zero values because each C=C bond is connected with the remaining part of the molecule via at least a single C−C bond. Besides, the same refers also to the negative component $D_{(+)i}^{(-4)}$ (but not to the positive one ($D_{(+)i}^{(+4)}$)). Finally, the second order member $D_{(+)i}^{(2)}$ evidently playes the decisive role in the formation of the total delocalization coefficient $D_{(+)i}$. This implies the ultimately positive sign of the latter and thereby an increased delocalization of any pair of $\pi-$electrons in polyenes as compared to that of an isolated C=C bond in accordance with Ref.[47]. Another important conclusion is about an essentially local origin of total delocalization coefficients of individual NCMOs and thereby of separate pairs of $\pi-$electrons. Moreover, this conclusion is straightforwardly extendable also to contributions of individual pairs of electrons ($\Delta E_I$) to the CEs of polyenes $\Delta E$ (see Eqs.(20) and (21)). The total CE of a certain polyene $\Delta E$ then accordingly consists of $N$ increments of semilocal nature representing the extents of participation of separate pairs of $\pi$-electrons. Again, the same CE of our polyene ($\Delta E$) is alternatively obtainable directly by summing up the partial delocalization coefficients of NCMOs in accordance with the left relation of Eq.(18). As discussed already, separate increments $d_{(+)i,(-)m}^{(k)}$ ($k=2$ and $k=4$) may be easily found by inspection. The overall procedure consists of the following steps: First, positive increments $\gamma^2/16$ of Eq.(22) should be ascribed to any pair of first-neighbouring C=C bonds. Second, positive contributions equal to $\gamma^4/64$ refer to all pairs of C=C bonds (both first- and second-neighbouring ones) inside all CP(3)s of the given polyene. Finally, negative increments should be found for each first-neighbouring pair of C=C bonds using the simple formula of Eq.(38). The total CE of the given polyene is then obtainable "on the back of an envelope" by summing up the above-enumerated increments in accordance with Eqs.(18) and (19). Therefore, we actually arrive at a simple additive scheme for the total CE of polyene in terms of essentially local increments representing particular aspects of delocalization of $\pi$-electrons.

## 4. Discussion of specific examples

The present Section is devoted to illustration of the above-discussed abstract results using the simplest polyenes as examples. Let us start with partial delocalization coefficients of NCMOs (Table 1). Second and fourth order increments to these coefficients are shown in this table for first five representatives of linear (I) and cross-conjugated (II) polyene chains (Fig. 1), where $N=2,3,...6$. [Note that the butadiene molecule ($N=2$) is the first member of both series under our interest, i.e. I($N=2$) coincides with II($N=2$)].

It is seen that uniform positive second order increments ($\gamma^2/16$) correspond to all pairs of first-neighbouring C=C bonds (denoted by $f$) of both linear (I) and cross-conjugated (II) chains, whatever the number of formally-double bonds ($N$). Moreover, coincidence of the increments $d_{(+)i,(-)m}^{(+2f)}$ for isomers I($N$) and II($N$) follows from comparison of the sixths columns of Table 1 referring to the same $N$ value. This result illustrates that of Eq.(22) and may be traced back to similar first neighbourhoods (adjacences) of C=C bonds of both chains.

Table 1. Separate increments to partial delocalization coefficients of NCMOs for simplest linear (I) and cross-conjugated (II) polyene chains

| Nr. comp. | $N$ | $(+)i$ | $(-)m$ | $f/s$ | $d^{(+2f)}_{(+)i,(-)m}$ | $d^{(+4f)}_{(+)i,(-)m}$ | $d^{(+4s)}_{(+)i,(-)m}$ | $d^{(-4f)}_{(+)i,(-)m}$ | $d^{(4)}_{(+)i,(-)m}$ |
|---|---|---|---|---|---|---|---|---|---|
| I | 2 | 1 | 2 | $f$ | $\gamma^2/16$ | - | - | $-3\gamma^4/256$ | $-3\gamma^4/256$ |
| I | 3 | 1 | 2 | $f$ | $\gamma^2/16$ | $4\gamma^4/256$ | - | $-6\gamma^4/256$ | $-2\gamma^4/256$ |
|   |   | 1 | 3 | $s$ | - | - | $4\gamma^4/256$ | - | $4\gamma^4/256$ |
| I | 4 | 1 | 2 | $f$ | $\gamma^2/16$ | $4\gamma^4/256$ | - | $-6\gamma^4/256$ | $-2\gamma^4/256$ |
|   |   | 2 | 3 | $f$ | $\gamma^2/16$ | $8\gamma^4/256$ | - | $-9\gamma^4/256$ | $-\gamma^4/256$ |
|   |   | 1 | 3 | $s$ | - | - | $4\gamma^4/256$ | - | $4\gamma^4/256$ |
| I | 5 | 1 | 2 | $f$ | $\gamma^2/16$ | $4\gamma^4/256$ | - | $-6\gamma^4/256$ | $-2\gamma^4/256$ |
|   |   | 2 | 3 | $f$ | $\gamma^2/16$ | $8\gamma^4/256$ | - | $-9\gamma^4/256$ | $-\gamma^4/256$ |
|   |   | 1 | 3 | $s$ | - | - | $4\gamma^4/256$ | - | $4\gamma^4/256$ |
|   |   | 2 | 4 | $s$ | - | - | $4\gamma^4/256$ | - | $4\gamma^4/256$ |
| I | 6 | 1 | 2 | $f$ | $\gamma^2/16$ | $4\gamma^4/256$ | - | $-6\gamma^4/256$ | $-2\gamma^4/256$ |
|   |   | 2 | 3 | $f$ | $\gamma^2/16$ | $8\gamma^4/256$ | - | $-9\gamma^4/256$ | $-\gamma^4/256$ |
|   |   | 3 | 4 | $f$ | $\gamma^2/16$ | $8\gamma^4/256$ | - | $-9\gamma^4/256$ | $-\gamma^4/256$ |
|   |   | 1 | 3 | $s$ | - | - | $4\gamma^4/256$ | - | $4\gamma^4/256$ |
|   |   | 2 | 4 | $s$ | - | - | $4\gamma^4/256$ | - | $4\gamma^4/256$ |
| II | 3 | 1 | 2 | $f$ | $\gamma^2/16$ | - | - | $-6\gamma^4/256$ | $-6\gamma^4/256$ |
|   |   | 1 | 3 | $s$ | - | - | - | - | - |
| II | 4 | 1 | 2 | $f$ | $\gamma^2/16$ | - | - | $-6\gamma^4/256$ | $-6\gamma^4/256$ |
|   |   | 2 | 3 | $f$ | $\gamma^2/16$ | - | - | $-9\gamma^4/256$ | $-9\gamma^4/256$ |
|   |   | 1 | 3 | $s$ | - | - | - | - | - |
| II | 5 | 1 | 2 | $f$ | $\gamma^2/16$ | - | - | $-6\gamma^4/256$ | $-6\gamma^4/256$ |
|   |   | 2 | 3 | $f$ | $\gamma^2/16$ | - | - | $-9\gamma^4/256$ | $-9\gamma^4/256$ |
|   |   | 1 | 3 | $s$ | - | - | - | - | - |
|   |   | 2 | 4 | $s$ | - | - | - | - | - |
| II | 6 | 1 | 2 | $f$ | $\gamma^2/16$ | - | - | $-6\gamma^4/256$ | $-6\gamma^4/256$ |
|   |   | 2 | 3 | $f$ | $\gamma^2/16$ | - | - | $-9\gamma^4/256$ | $-9\gamma^4/256$ |
|   |   | 3 | 4 | $f$ | $\gamma^2/16$ | - | - | $-9\gamma^4/256$ | $-9\gamma^4/256$ |
|   |   | 1 | 3 | $s$ | - | - | - | - | - |
|   |   | 2 | 4 | $s$ | - | - | - | - | - |

Analogous conclusions may be also drawn in respect of negative fourth order corrections $d^{(-4f)}_{(+)i,(-)m}$ defined by Eq.(38). Indeed, these corrections also refer to first-neighboring ($f$) pairs of C=C bonds only and are transferable when passing from a linear polyene to its cross-conjugated isomer for the same reason (i.e. similar adjacencies of C=C bonds). In contrast to second order corrections $d^{(+2f)}_{(+)i,(-)m}$, however, the fourth order ones ($d^{(-4f)}_{(+)i,(-)m}$) are no longer uniform for distinct pairs of C=C bonds inside the same chain. Moreover, the absolute values of the latter grow with increasing numbers of other first neighbours of the C=C bonds concerned in accordance of Eq.(38). For example, the absolute value of the correction $d^{(-4f)}_{(+)1,(-)2}$ grows when passing from butadiene ($N=2$) to more extended polyenes ($N>2$). Similarly, $\left|d^{(-4f)}_{(+)2,(-)3}\right|$ exceeds $\left|d^{(-4f)}_{(+)1,(-)2}\right|$ for $N=4,5,6$, etc.

By contrast, the positive fourth order corrections ($d^{(+4f)}_{(+)i,(-)m}$ and $d^{(+4s)}_{(+)i,(-)m}$) of Table 1 refer to both first- (*f*) and second-neighboring (*s*) pairs of C=C bonds. As discussed already, these corrections depend upon the presence of CP(3)s in the polyene concerned and thereby upon the type of conjugation. That is why the positive corrections are different for linear and cross-conconjugated isomers (I(*N*) and II(*N*)) and depend on the number of C=C bonds (*N*) in addition. The same then consequently refers also to total fourth order corrections ($d^{(4)}_{(+)i,(-)m}$) of the last column of Table 1. In this respect, individual polyenes of Fig. 1 deserve a comparative discussion.

Due to absence of CP(3)s in butadiene ($N=2$), the negative term $d^{(-4f)}_{(+)1,(-)2}$ is the only contribution to the relevant correction $d^{(4)}_{(+)1,(-)2}$, so that the latter also is a negative quantity. When passing to the linear hexatriene ($N=3$), a single CP(3) arises already which embraces all the three C=C bonds of this chain. Consequently, additional positive contributions $d^{(+4f)}_{(+)1,(-)2}$, $d^{(+4f)}_{(+)2,(-)3}$ and $d^{(+4s)}_{(+)1,(-)3}$ emerge along with the former negative increments. As a result, the absolute value of the total (negative) forth order correction $d^{(4)}_{(+)1,(-)2}$ ($d^{(4)}_{(+)2,(-)3}$) is lower for hexatriene as compared to that of butadiene in spite of the above-discussed growing absolute value of the negative increment $d^{(-4f)}_{(+)1,(-)2}$ with elongation of the chain. This implies higher ultimate values of partial delocalization coefficients $d_{(+)1,(-)2}$ and $d_{(+)2,(-)3}$ referring to first-neighboring C=C bonds in hexatriene vs. the analogous coefficient $d_{(+)1,(-)2}$ of butadiene [Note that a partial coefficient $d_{(+)i,(-)m}$ follows after summing up a large positive second order increment ($d^{(+2f)}_{(+)i,(-)m}$) and a relatively small fourth order correction ($d^{(4)}_{(+)i,(-)m}$)]. In addition, the pair of terminal C=C bonds ($C_1=C_4$ and $C_3=C_6$) of hexatriene is characterized by a positive partial delocalization coefficient ($d_{(+)1,(-)3}$) coinciding with the above-mentioned new term $d^{(+4s)}_{(+)1,(-)3}$. In summary, better conditions for delocalization of NCMOs (and thereby of electron pairs) may be concluded to be peculiar to the linear hexatriene as compared to butadiene and this outcome causes litle surprise.

Similar trends are preserved also when passing to the linear octatetraene ($N=4$). Presence of two CP(3)s in this chain gives rise to a two-fold positive increment $d^{(+4f)}_{(+)2,(-)3}$ referring to the "internal" pair of double bonds ($C_2=C_5$ and $C_3=C_8$) participating in both of these paths. As a result, the absolute value of the relevant fourth order contribution ($d^{(4)}_{(+)2,(-)3}$) is even lower. Meanwhile, the corrections $d^{(4)}_{(+)1,(-)2}$ and $d^{(4)}_{(+)3,(-)4}$ of octatetraene coincide with those of the linear hexatriene. This result illustrates the primary role of the nearest neighbourhood of the C=C bonds concerned in the formation of partial delocalization coefficients of NCMOs. It is then no surprise that the correction $d^{(4)}_{(+)1,(-)2}$ (referring to the terminal pairs of C=C bonds) exhibits no subsequent alterations also when passing to longer linear polyenes ($N=5,6$) in spite of growing numbers of CP(3)s. Therefore, extinction of the enhancement of the delocalization conditions is actually observed with further elongation of the linear polyene chain.

As opposed to the above-discussed linear chains, no CP(3)s are present in the cross-conjugated polyenes (II), whatever the number of C=C bonds *N*. This implies the negative terms $d^{(-4f)}_{(+)i,(-)m}$ to make the only contributions to total fourth order corrections $d^{(4)}_{(+)i,(-)m}$ for all cross-conjugated chains (II). As a result, the relevant corrections $d^{(4)}_{(+)i,(-)m}$ allways are of negative signs and of considerably higher absolute values in addition. It is evident that partial delocalization coefficients of NCMOs ($d_{(+)i,(-)m}$) then consequently take lower values as compared to those of the respective linear isomers. This distinction especially refers to the "internal" areas of the chains under comparison. In general, substantially worse delocalization conditions may be concluded to be peculiar to cross-conjugated polyenes as compared to their linear counterparts and the reason lies in the absence of CP(3)s in the former case. This result is in line with predictions of Ref.[2] based on numbers of the Dewar resonance structures.

Table 2. Second and fourth order increments to total delocalization coefficients of NCMOs and to conjugation energies of polyenes I-VIII

| Nr. comp. | $N$ | $(+)i$ | $D^{(2)}_{(+)i}$ | $E_{(2)I}$ | $E_{(2)}$ | $D^{(4)}_{(+)i}$ | $E_{(4)I}$ | $E_{(4)}$ |
|---|---|---|---|---|---|---|---|---|
| I | 2 | 1(2) | $\gamma^2/16$ | $\gamma^2/4$ | $\gamma^2/2$ | $-3\gamma^4/256$ | $-3\gamma^4/192$ | $-\gamma^4/32$ |
| I | 3 | 1(3) | $\gamma^2/16$ | $\gamma^2/4$ | $\gamma^2$ | $2\gamma^4/256$ | $2\gamma^4/192$ | 0 |
|   |   | 2 | $2\gamma^2/16$ | $2\gamma^2/4$ |  | $-4\gamma^4/256$ | $-4\gamma^4/192$ |  |
| I | 4 | 1(4) | $\gamma^2/16$ | $\gamma^2/4$ | $3\gamma^2/2$ | $2\gamma^4/256$ | $2\gamma^4/192$ | $\gamma^4/32$ |
|   |   | 2(3) | $2\gamma^2/16$ | $2\gamma^2/4$ |  | $\gamma^4/256$ | $\gamma^4/192$ |  |
| I | 5 | 1(5) | $\gamma^2/16$ | $\gamma^2/4$ | $2\gamma^2$ | $2\gamma^4/256$ | $2\gamma^4/192$ | $2\gamma^4/32$ |
|   |   | 2(4) | $2\gamma^2/16$ | $2\gamma^2/4$ |  | $\gamma^4/256$ | $\gamma^4/192$ |  |
|   |   | 3 | $2\gamma^2/16$ | $2\gamma^2/4$ |  | $6\gamma^4/256$ | $6\gamma^4/192$ |  |
| I | 6 | 1(6) | $\gamma^2/16$ | $\gamma^2/4$ | $5\gamma^2/2$ | $2\gamma^4/256$ | $2\gamma^4/192$ | $3\gamma^4/32$ |
|   |   | 2(5) | $2\gamma^2/16$ | $2\gamma^2/4$ |  | $\gamma^4/256$ | $\gamma^4/192$ |  |
|   |   | 3(4) | $2\gamma^2/16$ | $2\gamma^2/4$ |  | $6\gamma^4/256$ | $6\gamma^4/192$ |  |
| II | 3 | 1(3) | $\gamma^2/16$ | $\gamma^2/4$ | $\gamma^2$ | $-6\gamma^4/256$ | $-6\gamma^4/192$ | $-4\gamma^4/32$ |
|   |   | 2 | $2\gamma^2/16$ | $2\gamma^2/4$ |  | $-12\gamma^4/256$ | $-12\gamma^4/192$ |  |
| II | 4 | 1(4) | $\gamma^2/16$ | $\gamma^2/4$ | $3\gamma^2/2$ | $-6\gamma^4/256$ | $-6\gamma^4/192$ | $-7\gamma^4/32$ |
|   |   | 2(3) | $2\gamma^2/16$ | $2\gamma^2/4$ |  | $-15\gamma^4/256$ | $-15\gamma^4/192$ |  |
| II | 5 | 1(5) | $\gamma^2/16$ | $\gamma^2/4$ | $2\gamma^2$ | $-6\gamma^4/256$ | $-6\gamma^4/192$ | $-10\gamma^4/32$ |
|   |   | 2(4) | $2\gamma^2/16$ | $2\gamma^2/4$ |  | $-15\gamma^4/256$ | $-15\gamma^4/192$ |  |
|   |   | 3 | $2\gamma^2/16$ | $2\gamma^2/4$ |  | $-18\gamma^4/256$ | $-18\gamma^4/192$ |  |
| II | 6 | 1(6) | $\gamma^2/16$ | $\gamma^2/4$ | $5\gamma^2/2$ | $-6\gamma^4/256$ | $-6\gamma^4/192$ | $-13\gamma^4/32$ |
|   |   | 2(5) | $2\gamma^2/16$ | $2\gamma^2/4$ |  | $-15\gamma^4/256$ | $-15\gamma^4/192$ |  |
|   |   | 3(4) | $2\gamma^2/16$ | $2\gamma^2/4$ |  | $-18\gamma^4/256$ | $-18\gamma^4/192$ |  |
| III | 5 | 1 | $\gamma^2/16$ | $\gamma^2/4$ | $2\gamma^2$ | $2\gamma^4/256$ | $2\gamma^4/192$ | 0 |
|   |   | 2 | $2\gamma^2/16$ | $2\gamma^2/4$ |  | $6\gamma^4/256$ | $6\gamma^4/192$ |  |
|   |   | 3 | $3\gamma^2/16$ | $3\gamma^2/4$ |  | $-6\gamma^4/256$ | $-6\gamma^4/192$ |  |
|   |   | 4(5) | $\gamma^2/16$ | $\gamma^2/4$ |  | $-\gamma^4/256$ | $-\gamma^4/192$ |  |
| IV | 5 | 1 | $\gamma^2/16$ | $\gamma^2/4$ |  | $-6\gamma^4/256$ | $-6\gamma^4/192$ |  |
|  |  | 2 | $2\gamma^2/16$ | $2\gamma^2/4$ |  | $-7\gamma^4/256$ | $-7\gamma^4/192$ |  |
|  |  | 3 | $2\gamma^2/16$ | $2\gamma^2/4$ | $2\gamma^2$ | $-2\gamma^4/256$ | $-2\gamma^4/192$ | $-2\gamma^4/32$ |
|  |  | 4 | $2\gamma^2/16$ | $2\gamma^2/4$ |  | $\gamma^4/256$ | $\gamma^4/192$ |  |
|  |  | 5 | $\gamma^2/16$ | $\gamma^2/4$ |  | $2\gamma^4/256$ | $2\gamma^4/192$ |  |
| V | 5 | 1 | $\gamma^2/16$ | $\gamma^2/4$ |  | $-6\gamma^4/256$ | $-6\gamma^4/192$ |  |
|  |  | 2 | $2\gamma^2/16$ | $2\gamma^2/4$ |  | $-15\gamma^4/256$ | $-15\gamma^4/192$ | $-6\gamma^4/32$ |
|  |  | 3 | $2\gamma^2/16$ | $2\gamma^2/4$ | $2\gamma^2$ | $-10\gamma^4/256$ | $-10\gamma^4/192$ |  |
|  |  | 4 | $2\gamma^2/16$ | $2\gamma^2/4$ |  | $-7\gamma^4/256$ | $-7\gamma^4/192$ |  |
|  |  | 5 | $\gamma^2/16$ | $\gamma^2/4$ |  | $2\gamma^4/256$ | $2\gamma^4/192$ |  |
| VI | 5 | 1 | $\gamma^2/16$ | $\gamma^2/4$ |  | $7\gamma^4/256$ | $7\gamma^4/192$ |  |
|  |  | 2 | $3\gamma^2/16$ | $3\gamma^2/4$ |  | $-6\gamma^4/256$ | $-6\gamma^4/192$ |  |
|  |  | 3 | $2\gamma^2/16$ | $2\gamma^2/4$ | $2\gamma^2$ | $-2\gamma^4/256$ | $-2\gamma^4/192$ | 0 |
|  |  | 4 | $\gamma^2/16$ | $\gamma^2/4$ |  | $2\gamma^4/256$ | $2\gamma^4/192$ |  |
|  |  | 5 | $\gamma^2/16$ | $\gamma^2/4$ |  | $-\gamma^4/256$ | $-\gamma^4/192$ |  |
| VII | 5 | 1 | $2\gamma^2/16$ | $2\gamma^2/4$ |  | $-2\gamma^4/256$ | $-2\gamma^4/192$ |  |
|  |  | 2(4) | $2\gamma^2/16$ | $2\gamma^2/4$ | $2\gamma^2$ | $-7\gamma^4/256$ | $-7\gamma^4/192$ | $-2\gamma^4/32$ |
|  |  | 3(5) | $\gamma^2/16$ | $\gamma^2/4$ |  | $2\gamma^4/256$ | $2\gamma^4/192$ |  |
| VIII | 5 | 1(5) | $\gamma^2/16$ | $\gamma^2/4$ |  | $-6\gamma^4/256$ | $-6\gamma^4/192$ |  |
|  |  | 2(4) | $2\gamma^2/16$ | $2\gamma^2/4$ | $2\gamma^2$ | $-7\gamma^4/256$ | $-7\gamma^4/192$ | $-6\gamma^4/32$ |
|  |  | 3 | $2\gamma^2/16$ | $2\gamma^2/4$ |  | $-10\gamma^4/256$ | $-10\gamma^4/192$ |  |

    Comparison of total delocalization coefficients ($D_{(+)i}$) of separate NCMOs ($\psi_{(+)i}$) also supports the above-revealed distinction between linear (I) and cross-conjugated (II) polyene chains. Second and fourth order increments to these coefficients ($D^{(2)}_{(+)i}$ and $D^{(4)}_{(+)i}$) for $N=2,3,\ldots 6$ are shown in Table 2. It deserves recalling here that the sum $D^{(2)}_{(+)i}+D^{(4)}_{(+)i}$ serves as an approximation to $D_{(+)i}$ and reflects the overall extent of delocalization of the $I$th pair of $\pi-$electrons. Accordingly, the sum ($E_{(2)I}+E_{(4)I}$) represents the contribution of the same electron pair to the relevant CE. Total second and fourth order contributions to the latter ($E_{(2)}$ and $E_{(4)}$) also are exhibited nearby.

    The above-discussed transferability of second order increments to partial delocalization coefficients ($d^{(+2f)}_{(+)i,(-)m}$) for all pairs of first-neighbouring C=C bonds ensures a simple proportionality between $D^{(2)}_{(+)i}$ and the number of first neighbours of the $I$th C=C bond (see Eq.(23)), as well as uniform $D^{(2)}_{(+)i}$ values for C=C bonds characterized by the same nearest environment. Consequently, the second order increments ($D^{(2)}_{(+)i}$) to total delocalization coefficients take doubly lower values for terminal C=C bonds as compared to the internal ones for both linear (I) and cross-conjugated (II) polyenes. The same evidently refers to relative values of second order contributions ($E_{(2)I}$) to the total CE. Due to the primary importance of second order members ($D^{(2)}_{(+)i}$ and $E_{(2)I}$) in the formation both of total delocalization coefficients ($D_{(+)i}$) and of the total CE, considerably

suppressed delocalization of terminal electron pairs and thereby their reduced participation in conjugation may be regarded as the principal common feature of both polyene chains.

The relevant fourth order corrections ($D_{(+)i}^{(4)}$ and $E_{(4)I}$), in turn, contribute to formation of distinctions between individual C=C bonds and/or separate electron pairs of the same nearest environment if the second neighbourhoods are of different constitution (This equally refers to both terminal and internal C=C bonds). Accordingly, distinctions arise between linear and cross-conjugated chains due to the same reason. Thus, let us now dwell just on these corrections.

Let us start with linear polyenes I ($N=2,3…6$). It is seen that the fourth order corrections ($D_{(+)1}^{(4)}$ and $E_{(4)1}$) associated with a terminal C=C bond grow when passing from butadiene ($N=2$) to higher polyenes ($N=3,4…$). Similarly, the remaining corrections (referring to internal C=C bonds) also increase in their values with elongation of the chain. Finally, an analogous growth is observed when moving away from the terminal towards the middle of the latter. For example, the correction $D_{(+)3}^{(4)}$ (ascribed to the third C=C bond) exceeds considerably that of the second bond ($D_{(+)2}^{(4)}$) in polyenes I($N=5$) and I($N=6$). Thus, the above-concluded enhancement of delocalization conditions with elongation of the linear polyene (I) is additionally corroborated. Moreover, this trend is especially striking in the middle area of the chain, which actually ensures most of the CE of the given polyene. Besides, the bond length alternation (BLA) also is less pronounced in the middle areas of longer polyene chains [1,2,4,11].

The cross-conjugated polyenes (II) offer us an entirely opposite case in the same respects. Indeed, the corrections $D_{(+)1}^{(4)}$ and $E_{(4)1}$ take lower values when passing from butadiene to higher cross-conjugated chains. The same refers also to corrections ascribed to internal C=C bonds, viz. these exhibit an evident decrease with the increasing number of C=C bonds ($N$), as well as then moving towards the middle of the chain. Therefore, suppression of delocalization of $\pi-$electrons with elongation of the cross-conjugated polyene follows from the present analysis and this effect also is more conspicuous in the relevant central areas.

In summary, llinear and cross-conjugated polyenes are characterized by substantially different conditions of delocalization of individual pairs of $\pi$-electrons, and especially of those ascribed to the middle areas of these chains. This distinction is due to presence and absence of CP(3)s in the first and second case, respectively, and, consequently, it becomes more and more striking with elongation of the chains.

The above-discussed essentially local origin of both second and fourth order contributions to total delocalization coefficients of NCMOs (Section 3) allows us to expect analogous regularities in the relative extents of delocalization of separate electron pairs also for polyenes of less regular constitution, i.e. for those containing both linear and cross-conjugated fragments. To discuss this point, let us take the isomers III-VIII of decapentaene ($N=5$) as examples (Fig. 2). The relevant second and fourth order increments to total delocalization coefficients of NCMOs and to the CEs are shown in Table 2.

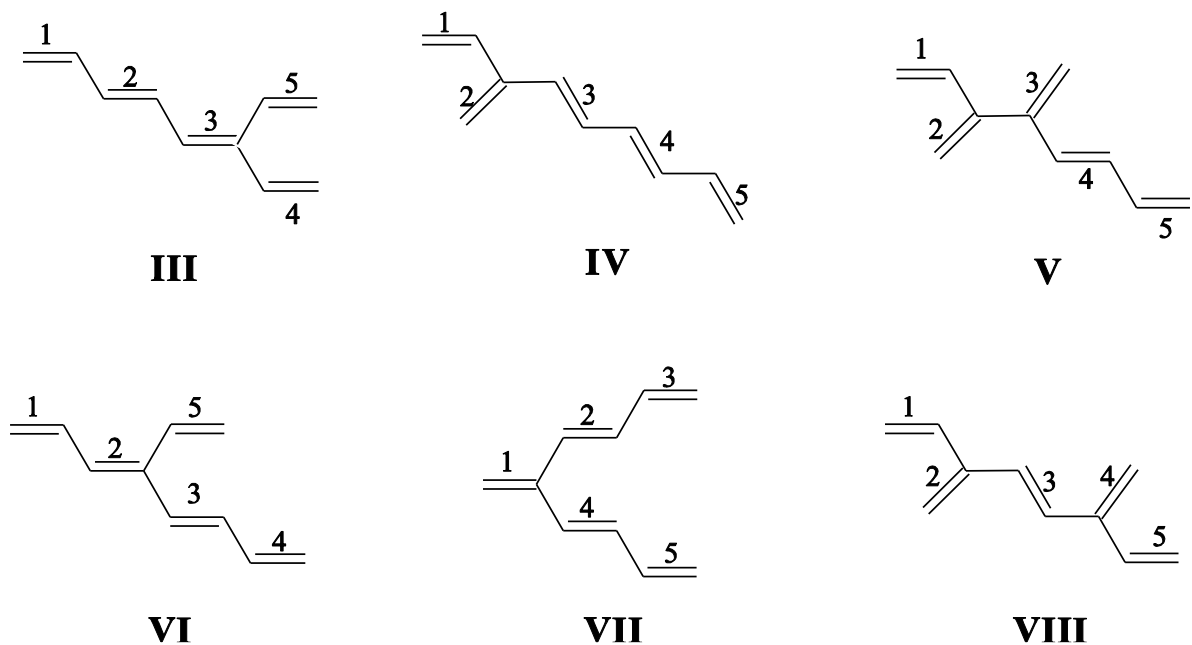

Figure 2: Selected isomers of decapentaene (III-VIII). Numberings of C=C bonds also are shown.

As already mentioned (Section 3), total second order energies ($E_{(2)}$) always are uniform for individual isomers of the same polyene due to coinciding numbers of CP(2)s and/or of C–C bonds. The isomers of decapentaene make no exception in this respect, viz. the eight hydrocarbons I($N$=5), II($N$=5) and III-VIII are characterized by the common $E_{(2)}$ value coinciding with $2\gamma^2$. The same evidently refers to second order increments ($D_{(+)}^{(2)}$) to the complete delocalization coefficients of these isomers. Meanwhile, contributions of separate electron pairs to these quantities generally are not uniform even for $k$=2. For example, contributions $D_{(+)i}^{(2)}$ and $E_{(2)I}$ take exceptionally high values for C=C bonds of isomers III and VI having three first neighbours in accordance with Eq.(23) [Contributions of the third and second C=C bond, respectively, are meant here]. This evidently indicates increased extents of delocalization of the relevant pairs of electrons (i.e. of those ascribed to the branching sites 3 and 2, respectively). At the same time, the isomers III and VI contain three more localized (terminal) electron pairs in contrast to only two pairs of this type present in the remaining isomers of decapentaene. The consequent reduction of the overall delocalization then compensates for its former increase due to branching so that the ultimate second order members of expansions are uniform both for the complete delocalization coefficients ($D_{(+)}^{(2)}$) and for the CEs ($E_{(2)}$). Thus, we arrive at a conclusion that quantitatively the same overall second order delocalization is distributed differently over separate C=C bonds within individual isomers. In particular, this distribution is less homogeneous in the branched isomers of decapentaene III and VI as compared to the remaining ones.

The fourth order contributions both to delocalization of $\pi$-electrons and to the CEs of isomers III-VIII also obey the above-established rules. First of all, couples of isomers III and VI, IV and VII, as well as V and VIII, are characterized by uniform values of total corrections $D_{(+)}^{(4)}$ and $E_{(4)}$, i.e. we have to do here with couples of isoenergetic systems to within fourth order terms of our power series inclusive. Nevertheless, the relevant contributions of separate pairs of $\pi$-electrons and/or of C=C bonds exhibit considerable differences in all isomers concerned. Thus, quantitatively the same total fourth order delocalization inside isoenergetic couples also is distributed differently over separate C=C bonds and/or electron pairs and this distribution seems to be governed by respective local structures in addition. Indeed, C=C bonds belonging to cross-conjugated fragments of isomers III-VIII generally are characterized by negative corrections $D^{(4)}_{(+)i}$ and $E_{(4)I}$, e.g. the bonds under numbers 3,4,5 and 1,2,3 of isomers III and IV, respectively, as well as 2,3,5 and 1,2,4 of isomers VI and VII, etc. Meanwhile, the same corrections prove to be increased (or even take positive values) for C=C bonds of linear fragments, especially for those participating in two CP(3)s. For example, higher positive values of corrections $D^{(4)}_{(+)2}$ and $D^{(4)}_{(+)1}$ of isomers III and VI, respectively, may be traced back to participation of the C=C bonds concerned in two CP(3)s. The same reason seems to underly the sufficiently high value of the increment $D^{(4)}_{(+)1}$ of isomer VII in spite of pertinence of the first C=C bond to a cross-conjugated fragment. Analogously, a higher value of $D^{(4)}_{(+)2}$ of isomer VIII as compared to the same correction of its counterpart V also is largely due to participation of the second C=C bond in a CP(3) in the former decapentaene but not in the latter one.

### 5. Conclusions

Application of the non-canonical method of MOs offers us an alternative viewpoint of conjugation in acyclic polyenes. Accordingly, the effect of conjugation of $N$ double (C=C) bonds manifests itself in a weak and rather local delocalization of respective $N$ initially-localized pairs of $\pi$–electrons. The relevant conjugation energy (CE) then coincides with the total delocalization energy of all these pairs. Analogously, relative stability of a certain polyene depends upon the extent of the underlying delocalization.

The above-formulated interrelation between the extents of delocalization and stabilization is valid also for each pair of $\pi$-electrons separately. Indeed, the total CE of any acyclic polyene is expressible as a sum of $N$ contributions, each of them representing an individual pair of $\pi$-electrons and directly related to the total delocalization coefficient of the respective single non-canonical (localized) MO. This implies that the more delocalized the $I$th pair of $\pi-$ electrons becomes, the higher its contribution to the total CE gets and vice versa.

In addition, the total delocalization coefficient of the $I$th non-canonical MO is shown to determine the extent of reduction in the "internal" order of the parent ($I$th) formally-double (C=C) bond when building up the polyene concerned. In other words, weaker formally-double bonds correspond to more delocalized pairs of $\pi$-electrons and vice versa. It deserves mention in this context that local manifestations of the conjugation effect (such as lengthening of individual formally-double bonds and shortening of initially-single ones) also are likely to be related to the relevant local delocalization.

The extents of delocalization of individual pairs of $\pi$-electrons, in turn, prove to be determined by two structural factors of local nature. First, delocalization of a certain pair is induced by participation of the respective parent C=C bond in both two- and three-membered conjugated paths. Evidently, constitutions only of the first and of the second neighbourhoods of the parent bond are of importance here. Second, each pair of $\pi$-electrons is undergoing a definite repulsion from its first neighbours that depends also on the relevant number of second neighbours and inhibits the delocalization concerned. Consequently, a local relation may be concluded to exist between constitution of the nearest environment of the $I$th C=C bond, delocalization pattern of the $I$th pair of $\pi-$ electrons and contribution of this pair to the total CE of the given polyene.

For illustration of the above-summarized general results, two polyenes of regular constitution have been taken as the principal examples, viz. the linear chain (I) and the cross-conjugated one (II). It turned out that negative (suppressive) increments to delocalization take uniform values for both chains under comparison due to coinciding numbers of both first and second neighbours of individual C=C bonds. Meanwhile, the relevant positive contributions are substantially different. The point is that the linear conjugation of C=C bonds allows participation of the latter in a certain number of three-membered conjugated paths (CP(3)s) in addition to the two-membered ones (CP(2)s) equally inherent in both chains. Moreover, emergence of a CP(3) (e.g. I–J–K) not only provides an increase in the short-range delocalization caused by CP(2)s (e.g. of the *I*th pair of $\pi-$electrons over the *J*th C=C bond), but also it gives birth to a long-range delocalization (viz. of the *I*th pair over the *K*th C=C bond and vice versa). In summary, substantially better delocalization conditions may be concluded to be peculiar to linear polyenes (I) as compared to their cross-conjugated counterparts (II). Since the internal C=C bonds always are embraced by a higher number of CP(3)s as compared to the terminal ones, distinction between the chains I and II is especially striking in the middle areas of the latter. Finally, an enhancement (deterioration) of the delocalization conditions is observed with elongation of a linear (cross-conjugated) chain.

Polyenes of irregular constitution (e.g. those containing both linear and cross-conjugated fragments) also illustrate the local origin of delocalization. Indeed, delocalization of electron pair(s) proves to be generally increased and supressed, if the respective parent C=C bond(s) belong(s) to linear and to cross-conjugated fragments(s), respectively. Higher (lower) contributions of these pairs to the total CEs of polyenes also are among the conclusions.